\documentclass[12pt]{article}
\usepackage{amssymb}
\usepackage{epsfig}
\usepackage{float}
\usepackage{graphicx}
\usepackage{amsmath}
\newcommand{\nua}[1]{\ensuremath{\rlap
           {\kern-2.5pt\ensuremath
           {\overset{\scriptscriptstyle(-)}{\phantom{\nu}}}}
           {\ensuremath{{\nu}_{#1}}}}}

\begin{document}
\begin{center}
{\bf Some comments on high precision study of  neutrino oscillations}

\end{center}
\begin{center}
S. M. Bilenky
\end{center}
\begin{center}
{\em  Joint Institute for Nuclear Research, Dubna, R-141980,
Russia\\}
{\em TRIUMF
4004 Wesbrook Mall,
Vancouver BC, V6T 2A3
Canada\\}
\end{center}

\begin{abstract}
 I discuss  some problems connected with the high precision study of neutrino oscillations. In the general case of $n$-neutrino mixing I  derive  a convenient expression for transition probability in which only independent  terms (and mass-squared differences) enter. For three-neutrino mixing I discuss a problem of a definition of a large (atmospheric) neutrino mass-squared difference. I comment also possibilities to reveal the character of neutrino mass spectrum in future reactor neutrino experiments.

\end{abstract}

\section{Introduction}
The observation of neutrino oscillations  in the atmospheric
Super-Kamiokande \cite{Fukuda:1998mi}, solar SNO \cite{Ahmad:2002jz}, reactor KamLAND \cite{Araki:2004mb} and solar neutrino oscillation experiments \cite{Cleveland:1998nv,Altmann:2005ix,Abdurashitov:2002nt} is one of the most important recent discovery in the particle physics.

Small neutrino masses, many orders of magnitude smaller than masses of other fundamental fermions, are  an evidence of a beyond the Standard Model physics. One of the plausible scenario which allow to explain the smallness of neutrino masses is based on the assumption that small (Majorana) neutrino masses are generated by the lepton-number violating dimension five effective Lagrangian \cite{Weinberg:1979sa}. In this case neutrino masses are suppressed with respect to masses of leptons and quarks by the ratio of the electroweak scale $v=(\sqrt{2}G_{F})^{-1/2}\simeq 246$ GeV and a scale $\Lambda\gg v$ of a new lepton number-violating physics.

Neutrino oscillation data can be  described by the three-neutrino mixing
\begin{equation}\label{mix}
\nu_{lL}(x)=\sum^{3}_{i=1}U_{li}~\nu_{iL}(x).\quad (l=e,\mu,\tau)
\end{equation}
Here $\nu_{i}(x)$ is the field of neutrinos (Dirac or Majorana) with mass $m_{i}$ and $U$ is the unitary $3\times3$ PMNS \cite{Pontecorvo:1957cp,Pontecorvo:1957qd,Maki:1962mu} mixing matrix.

In the framework of the three-neutrino mixing neutrino oscillations are characterized by two neutrino mass-squared differences $\Delta m^{2}_{23}$ and $\Delta m^{2}_{12}$, three mixing angles $\theta_{12}$, $\theta_{23}$, $\theta_{13}$ and one $CP$ phase $\delta$. From the analysis of the data of neutrino oscillation experiments it was established that $\Delta m^{2}_{12}\ll \Delta m^{2}_{23}$, mixing angles $\theta_{23}$
and $\theta_{12}$ are large and  mixing angle  $\theta_{13}$  is small. The first information about the angle $\theta_{13}$ was obtained from the reactor CHOOZ experiment \cite{Apollonio:1999ae} in which only the upper bound $\sin^{2}2\theta_{13}\leq 1\cdot 10^{-1}$ was found.

First data of neutrino oscillation experiments were described by expressions for neutrino transition probabilities in the leading approximation which was based on the assumption that $\sin^{2}\theta_{13}=0$. In this approximation  oscillations in atmospheric and KamLAND (solar) regions are decoupled (see \cite{Bilenky:1998dt}): in the atmospheric region (atmospheric and long-baseline accelerator neutrino oscillation experiments) neutrino oscillations are two-neutrino $\nu_{\mu}\rightleftarrows\nu_{\tau} $ oscillations,  in the solar region (the reactor KamLAND experiment) neutrino oscillations are $\bar\nu_{e}\rightleftarrows\bar\nu_{\mu,\tau} $ oscillations. From analysis of the atmospheric and long-baseline accelerator  oscillation experiments parameters  $\Delta m^{2}_{23}$
and $\sin^{2}2\theta_{23}$ were determined. From analysis of the data of the KamLAND and solar experiments another two neutrino oscillation parameters $\Delta m^{2}_{12}$
and $\sin^{2}2\theta_{12}$ were inferred. In the leading approximation the character of the neutrino mass spectrum and such important  effect of the three-neutrino mixing as $CP$ violation in the lepton sector can not be revealed.

With the measurement of the mixing angle  $\theta_{13}$ in the reactor Daya Bay \cite{An:2013zwz}, RENO \cite{Ahn:2012nd} and Double CHOOZ \cite{Abe:2013sxa} experiments the situation with the study of neutrino oscillations drastically changed. The investigation of neutrino oscillations entered into  high precision era, era of measurements of small, beyond the leading approximation effects which could allow to determine the character of the neutrino mass spectrum and  to measure $CP$ phase $\delta$.

In this paper for the general case of the $n$-neutrino mixing we will derive a convenient expression for the neutrino transition probability in vacuum in which  only independent terms (and mass-squared differences) enter.

In different papers large (atmospheric) neutrino mass-squared difference is determined differently. Difference between different definitions is small (a few \%) but in the era of precision measurements apparently it is desirable to have one unified definition. The expression for transition probability we will present here provides  natural framework for introduction of two independent neutrino mass-squared differences in the case of the three-neutrino mixing.

Determination of the character of the neutrino mass spectrum is one of the major aim of future reactor neutrino experiments JUNO \cite{Li:2014qca} and RENO-50 \cite{Kim:2014rfa}. On the basis of the proposed expression for the transition probability I will comment this possibility.

\section{General expression for neutrino  transition probability in vacuum}
For the general case of the neutrino mixing
\begin{equation}\label{genmix}
\nu_{\alpha L}(x)=\sum^{3+n_{s}}_{i=1}U_{\alpha i}~\nu_{iL}(x)\quad (\alpha=e, \mu, \tau, s_{1},...s_{n_{s}})
\end{equation}
we will derive here an expression for $\nu_{\alpha}\to \nu_{\alpha'}$ transition probability alternative to the standard one. Here $n_{s}$ is the number of sterile neutrino fields, $U$ is an unitary $(3+n_{s})\times (3+n_{s})$ mixing matrix, $\nu_i(x)$ is the field of neutrino with mass $m_{i}$.

From (\ref{genmix}) and Heisenberg uncertainty relation it follows that normalized
states of flavor $\nu_{e}, \nu_{\mu}, \nu_{\tau}$ and sterile $\nu_{s_{1}}, \nu_{s_{2}},...$  neutrinos are described by {\em coherent superpositions} of  the states of  neutrinos with definite masses (see, for example, \cite{Bilenky:1987ty,Bilenky:1998dt,Bilenky:2001yh})
\begin{equation}\label{mixedstate}
|\nu_\alpha\rangle
=
\sum^{n}_{i=1} U_{\alpha i}^* \,~ |\nu_i\rangle.
\end{equation}
Here $|\nu_i\rangle$ is the state of the left-handed neutrino with mass $m_{i}$, momentum $\vec{p}$ and energy
$E_{i}=\sqrt{p^{2}+m^{2}_{i}}\simeq E+ \frac{m^{2}_{i}}{2E}$ ($E=p$ is the energy of neutrino at $m_{i}\to 0$).

If at $t=0$ flavor neutrino $\nu_\alpha$ is produced, at the time $t$ we have
\begin{equation}\label{mixedstate2}
|\nu_\alpha\rangle_{t}= \sum_{\alpha'}|\nu_{\alpha'}\rangle\langle\nu_{\alpha'}|
e^{-iH_{0}t}|\nu_\alpha\rangle=
 \sum_{\alpha'}|\nu_{\alpha'}\rangle (\sum_{i}U_{\alpha'i},
e^{-iE_{i}t}U^{*}_{\alpha i})
\end{equation}
where $H_{0}$ is the free Hamiltonian.

From (\ref{mixedstate2}) for the normalized probability of the  $\nu_\alpha\to \nu_{\alpha'}$ transition  we find the following expression
\begin{equation}\label{standard}
P( \nu_\alpha\to \nu_{\alpha'})=|\sum_{i}U_{\alpha'i},
e^{-iE_{i}t}U^{*}_{\alpha i}|^{2}=\sum_{i}|U_{\alpha' i}|^{2}  |U_{\alpha i}|^{2}+ 2~\sum_{i>k}\mathrm{Re}(U_{\alpha'i}
U^{*}_{\alpha i}U^{*}_{\alpha'k}
U_{\alpha k}  e^{-2i\Delta _{ki}}).
\end{equation}
Here
\begin{equation}\label{phase}
\Delta _{ki}= \frac{\Delta m^{2}_{ki}L}{4E},
\end{equation}
where $\Delta m^{2}_{ki}= m^{2}_{k}- m^{2}_{i}$ and $L\simeq t$ is the neutrino source-detector distance.

Taking into account the unitarity of the mixing matrix $U$ for the first term of the probability (\ref{standard}) we have
\begin{equation}\label{standard1}
 \sum_{i}|U_{\alpha' i}|^{2}  |U_{\alpha i}|^{2}=\delta _{\alpha' \alpha}- 2~\sum_{i>k}\mathrm{Re}(U_{\alpha'i}
U^{*}_{\alpha i}U^{*}_{\alpha'k}
U_{\alpha k}).
\end{equation}
From (\ref{standard}) and (\ref{standard1}) for the $\nua{\alpha}\to \nua{\alpha'}$ transition probability  we obtain the following {\em standard expression} (see \cite{GonzalezGarcia:2007ib,Giunti:2007ry,Bilenky:2010zza})
\begin{eqnarray}\label{standard4}
P(\bar \nu_\alpha\to \bar\nu_{\alpha'})&=&\delta _{\alpha' \alpha}
-4~\sum_{i>k}\mathrm{Re}~(U_{\alpha'i}
U^{*}_{\alpha i}U^{*}_{\alpha'k}
U_{\alpha k})\sin^{2}\Delta _{ki}\nonumber\\
&\pm& 2~\sum_{i>k}\mathrm{Im}~(U_{\alpha'i}
U^{*}_{\alpha i}U^{*}_{\alpha'k}
U_{\alpha k})\sin 2\Delta _{ki}.
\end{eqnarray}
Let us stress that not all quantities in (\ref{standard4}) are independent. For example, in the case of the three-neutrino mixing three mass-squared differences (\ref{standard4}) are connected by the relation  $\Delta m^{2}_{13}=\Delta m^{2}_{12} +\Delta m^{2}_{23}$. For $\alpha'\neq \alpha$ the quantities in the last term
of  (\ref{standard4}) are connected by the relations $\mathrm{Im}~(U_{\alpha'2}U^{*}_{\alpha 2}U^{*}_{\alpha'1} U_{\alpha 1})=\mathrm{Im}~(U_{\alpha'3}U^{*}_{\alpha 3}U^{*}_{\alpha'2} U_{\alpha 2})=-\mathrm{Im}~(U_{\alpha'3}U^{*}_{\alpha 3}U^{*}_{\alpha'1} U_{\alpha 1})$ which follow from the unitarity of the mixing matrix (see \cite{Giunti:2007ry,Bilenky:2010zza}).

We will obtain here a simple expression for  the neutrino transition probability in vacuum in which
\begin{itemize}
  \item we will take into account that there is  one arbitrary common phase in the transition amplitude,
  \item we will use  the unitarity of the mixing matrix in the transition amplitude.
\end{itemize}

We have
\begin{eqnarray}\label{Genexp1}
  P(\nu_{\alpha}\to \nu_{\alpha'})&=&|\sum_{i}U_{\alpha'i},
e^{-i(E_{i}-E_{p})t}U^{*}_{\alpha i}|^{2}=|\delta_{\alpha'\alpha}+
\sum_{i\neq p}U_{\alpha'i}~(e^{-2i\Delta _{pi}}-1)~
U^{*}_{\alpha i}|^{2}\nonumber\\&=&|\delta_{\alpha'\alpha}-2i\sum_{i\neq p}U_{\alpha'i}~
U^{*}_{\alpha i}e^{-i\Delta_{pi}}\sin\Delta_{pi}|^{2},
\end{eqnarray}
where $p$ is an arbitrary fixed index.

From (\ref{Genexp1}) we find
\begin{eqnarray}
&&P(\nu_{\alpha}\to \nu_{\alpha'})=\delta_{\alpha'\alpha}
-4\sum_{i\neq p}|U_{\alpha i}|^{2}(\delta_{\alpha' \alpha } - |U_{\alpha' i}|^{2})\sin^{2}\Delta_{pi}
\nonumber\\
&&+8~\sum_{i>k;i,k\neq p}\mathrm{Re}(U_{\alpha' i}U^{*}_{\alpha i}U_{\alpha'
k}^{*}U_{\alpha k}~e^{-i(\Delta_{pi}-\Delta_{pk})})\sin\Delta_{pi}\sin\Delta_{pk}.
\label{Genexp3}
\end{eqnarray}
Finally, we obtain the following general expression for $\nua{\alpha}\to \nua{\alpha'}$ transition probability
\cite{Bilenky:2012zp}
\begin{eqnarray}
P(\nua{\alpha}\to \nua{\alpha'})
=\delta_{\alpha' \alpha }
-4\sum_{i\neq p}|U_{\alpha i}|^{2}(\delta_{\alpha' \alpha } - |U_{\alpha' i}|^{2})\sin^{2}\Delta_{pi}\nonumber\\
+8~\sum_{i>k;i,k\neq p }\mathrm{Re}~(U_{\alpha' i}U^{*}_{\alpha i}U^{*}_{\alpha'
k}U_{\alpha k})\cos(\Delta_{pi}-\Delta_{pk})\sin\Delta_{pi}\sin\Delta_{pk}\nonumber\\
\pm 8~\sum_{i>k;i,k\neq p}\mathrm{Im}~(U_{\alpha' i}U^{*}_{\alpha i}U^{*}_{\alpha'
k}U_{\alpha k})\sin(\Delta_{pi}-\Delta_{pk})\sin\Delta_{pi}\sin\Delta_{pk},
\label{Genexp4}
\end{eqnarray}
where sign + (-) refers to $\nu_{\alpha}\to \nu_{\alpha'}$ ($\bar\nu_{\alpha}\to \bar\nu_{\alpha'}$) transition.

In (\ref{Genexp4}) only independent terms (and mass-squared differences) enter. For example, for the three-neutrino mixing there are only two independent mass-squared differences and one $i>k$ term in the transition probability (because $i,k\neq p$).

\section{Three-neutrino oscillations}
\subsection{Atmospheric neutrino mass-squared difference? Flavor neutrino transition probability}

From  analysis of the neutrino oscillation data it follows that one mass-squared difference
(atmospheric) is much larger than the other one (solar). Two neutrino mass spectra are possible in such a situation \footnote{Usually neutrinos with small mass-squared difference are called $\nu_{1}$ and $\nu_{2}$. It is assumed also that $m_{2}>m_{1}$, i.e. that  $\Delta m^{2}_{12}>0$.}
\begin{enumerate}
  \item Neutrino spectrum  with small mass-squared difference between lightest neutrinos (Normal spectrum, NS)

  $m_{1}<m_{2}<m_{3},\quad \Delta m_{12}^{2}\ll \Delta m_{23}^{2}$

   \item  Neutrino spectrum  with small mass-squared difference between heaviest neutrinos (Inverted spectrum, IS)

$m_{3}<m_{1}<m_{2},\quad \Delta m_{12}^{2}\ll |\Delta m_{13}^{2}|$

\end{enumerate}
There are only two possibilities to introduce small (solar) $\Delta m_{S}^{2}$ and large (atmospheric) $\Delta m_{A}^{2}$ mass-squared differences in the framework of the approach we are advocating \footnote{Notice that the first option corresponds to extraction  of the phase connected with the intermediate neutrino mass
in the expression (\ref{Genexp1})
and second option corresponds to extraction of the phase connected with last mass.
}.
\begin{enumerate}
  \item
\begin{equation}\label{NNS}
\mathrm{NS.}~~~\Delta m_{21}^{2}=-\Delta m_{S}^{2},\quad \Delta m_{23}^{2}=\Delta m_{A}^{2} ~~(p=2)
\end{equation}

 \begin{equation}\label{IIS}
\mathrm{IS.}~~~ \Delta m_{12}^{2}=\Delta m_{S}^{2},\quad \Delta m_{13}^{2}=-\Delta m_{A}^{2}~~ (p=1)
 \end{equation}

 \item
\begin{equation}\label{NNNS}
\mathrm{NS.}~~~ \Delta m_{12}^{2}=\Delta m_{S}^{2},\quad \Delta m_{13}^{2}=\Delta m_{A}^{2}~~(p=1)
\end{equation}

\begin{equation}\label{IIIS}
\mathrm{IS.}~~~ \Delta m_{21}^{2}=-\Delta m_{S}^{2},\quad \Delta m_{23}^{2}=-\Delta m_{A}^{2}~~ (p=2)
\end{equation}

\end{enumerate}

In all papers on neutrino oscillations  mixing angles, $CP$ phase and solar mass-squared difference are determined in same way. However, atmospheric mass-squared difference in
different papers is determined differently. For example, (in terms of parameters introduced in 1.)
\begin{enumerate}
  \item The Bari group determines  large neutrino mass-squared difference  as follows (see \cite{Capozzi:2013psa})
\begin{equation}\label{3numix4}
\Delta m^{2}=\frac{1}{2}| \Delta m_{13}^{2}+\Delta m_{23}^{2}|=\Delta m_{A}^{2}+\frac{1}{2}\Delta m_{S}^{2}
\end{equation}
  \item The NuFit group determines
the atmospheric mass-squared difference as in 2. (see \cite{Gonzalez-Garcia:2014bfa})
\begin{equation}\label{3numix5}
\Delta m_{13}^{2}=\Delta m_{A}^{2}+\Delta m_{S}^{2},~~~(NS),~~~ \Delta m_{23}^{2}=-(\Delta m_{A}^{2}+\Delta m_{S}^{2})~~~(IS).
\end{equation}
  \item In the T2K paper \cite{Abe:2014ugx} the atmospheric mass-squared difference is determined as in 1.
  \item In the MINOS paper \cite{Adamson:2014vgd} large mass-squared difference is determined as $|\Delta m_{23}^{2}|$ for both mass spectra. It is obvious, however,
that $\Delta m_{23}^{2}$ for NS and $|\Delta m_{23}^{2}|$ for IS are different quantities.
\end{enumerate}
The difference between different "atmospheric neutrino mass-squared differences" is a few \%. It is determined by the ratio  $\frac{ \Delta m_{S}^{2}}{ \Delta m_{A}^{2}}\simeq 3\cdot 10^{-2}$ and can not be neglected in the precision era. Apparently one definition is desirable.

We will choose here the option 1. From (\ref{Genexp4}) in the case of normal and inverted neutrino mass spectra we have respectively
\begin{eqnarray}
&&P^{NS}(\nua{l}\to \nua{l'})
=\delta_{l' l }
-4|U_{l 3}|^{2}(\delta_{l' l} - |U_{l' 3}|^{2})\sin^{2}\Delta_{A}\nonumber\\&&-4|U_{l 1}|^{2}(\delta_{l' l} - |U_{l' 1}|^{2})\sin^{2}\Delta_{S}
-8~\mathrm{Re}~(U_{l' 3}U^{*}_{l 3}U^{*}_{l'
1}U_{l 1})\cos(\Delta_{A}+\Delta_{S})\sin\Delta_{A}\sin\Delta_{S}\nonumber\\
&&\mp 8~\mathrm{Im}~(U_{l' 3}U^{*}_{l 3}U^{*}_{l'
1}U_{l 1})\sin(\Delta_{A}+\Delta_{S})\sin\Delta_{A}\sin\Delta_{S},
\label{Genexp5}
\end{eqnarray}
and
\begin{eqnarray}
&&P^{IS}(\nua{l}\to \nua{l'})
=\delta_{l' l }
-4|U_{l 3}|^{2}(\delta_{l' l } - |U_{l' 3}|^{2})\sin^{2}\Delta_{A}\nonumber\\&&-4|U_{l 2}|^{2}(\delta_{l' l} - |U_{l' 2}|^{2})\sin^{2}\Delta_{S}
-8~\mathrm{Re}~(U_{l' 3}U^{*}_{l 3}U^{*}_{l'
2}U_{l 2})\cos(\Delta_{A}+\Delta_{S})\sin\Delta_{A}\sin\Delta_{S}\nonumber\\
&&\pm 8~\mathrm{Im}~(U_{l' 3}U^{*}_{l 3}U^{*}_{l'
2}U_{l 2})\sin(\Delta_{A}+\Delta_{S})\sin\Delta_{A}\sin\Delta_{S}.
\label{Genexp6}
\end{eqnarray}
Here
\begin{equation}\label{Delta}
\Delta_{A,S}=\frac{\Delta m^{2}_{A,S}L}{4E},
\end{equation}
Thus, transition probabilities depend on "extreme values" of the elements of neutrino mixing matrix: $U_{l'1(3)}$ and $U_{l1(3)}$ in the NS case ($p=2$) and
$U_{l'2(3)}$ and $U_{l2(3)}$ in the IS case ($p=1$). Difference in signs of the last terms of (\ref{Genexp5}) and (\ref{Genexp6}) is connected with signs in (\ref{NNS}) and (\ref{IIS}).

If $CP$ is violated in the lepton sector in this case we have
\begin{equation}\label{1CP}
P(\nu_{l}\to \nu_{l'})\neq P(\bar\nu_{l}\to \bar\nu_{l'})\quad (l'\neq l)
\end{equation}
 Let us determine the $CP$ asymmetry
\begin{equation}\label{CPasymm}
A^{CP}_{l' l }=P(\nu_{l}\to \nu_{l'})-P(\bar\nu_{l}\to \bar\nu_{l'})
\end{equation}
The $CP$ asymmetry satisfies the following general conditions
\begin{equation}\label{CPasymm1}
A^{CP}_{l' l }=-A^{CP}_{l l'}.
\end{equation}
and
\begin{equation}\label{CPasymm2}
\sum_{l'}A^{CP}_{l' l }=0.
\end{equation}
The first condition follows from the relation
\begin{equation}\label{CPasymm3}
P(\nu_{l}\to \nu_{l'})= P(\bar\nu_{l'}\to \bar\nu_{l})
\end{equation}
which is a consequence of  the $CPT$ invariance. The second condition follows from the conservation of the probability
\begin{equation}\label{CPasymm4}
 \sum_{l'}P(\nu_{l}\to \nu_{l'})=\sum_{l'}P(\bar\nu_{l}\to \bar\nu_{l'})=1.
\end{equation}
From (\ref{CPasymm1}) and (\ref{CPasymm2}) it follows that in the case of the three-neutrino mixing
$CP$ asymmetries in different flavor channels are connected by the following relations \cite{Bilenky:1981hf}
\begin{equation}\label{CPasymm5}
A^{CP}_{\mu e }=A^{CP}_{e \tau}=-A^{CP}_{\mu \tau }.
\end{equation}
From (\ref{Genexp5}) in the case NS we have
\begin{equation}\label{CPNH}
A^{CP}_{l' l }=-16~\mathrm{Im}~U_{l' 3}U^{*}_{l 3}U^{*}_{l'
1}U_{l 1}\sin(\Delta_{A}+\Delta_{S})\sin\Delta_{A}\sin\Delta_{S}.
\end{equation}
For IS from (\ref{Genexp6}) we find
\begin{equation}\label{CPIH}
A^{CP}_{l' l }=16~\mathrm{Im}~U_{l' 3}U^{*}_{l 3}U^{*}_{l'
2}U_{l 2}\sin(\Delta_{A}+\Delta_{S})\sin\Delta_{A}\sin\Delta_{S}.
\end{equation}
In the next subsections we will present expressions for transition probabilities which are of experimental interest. For that we will use the standard parametrization of the PMNS mixing matrix
\begin{eqnarray}
U=\left(\begin{array}{ccc}c_{13}c_{12}&c_{13}s_{12}&s_{13}e^{-i\delta}\\
-c_{23}s_{12}-s_{23}c_{12}s_{13}e^{i\delta}&
c_{23}c_{12}-s_{23}s_{12}s_{13}e^{i\delta}&c_{13}s_{23}\\
s_{23}s_{12}-c_{23}c_{12}s_{13}e^{i\delta}&
-s_{23}c_{12}-c_{23}s_{12}s_{13}e^{i\delta}&c_{13}c_{23}
\end{array}\right).
\label{unitmixU1}
\end{eqnarray}
Here $c_{12}=\cos\theta_{12}$,  $s_{12}=\sin\theta_{12}$ etc.

\subsection{$\bar\nu_{e}\to \bar\nu_{e}$ survival probability}
Expressions for the three-neutrino $\bar\nu_{e}$ survival probabilities are important for
analysis of the data of the reactor neutrino experiments.
From (\ref{Genexp5}) and  (\ref{Genexp6}) for normal and inverted mass ordering we have respectively
\begin{eqnarray}
&&P^{\mathrm{NS}}(\bar\nu_{e}\to \bar\nu_{e})=1- 4~|U_{e
3}|^{2}(1-|U_{e3}|^{2})~ \sin^{2}\Delta_{A}
\nonumber\\
&&- 4~|U_{e 1}|^{2}(1-|U_{e 1}|^{2})~ \sin^{2}\Delta_{S}
\nonumber\\
&&-8~|U_{e 3}|^{2}|U_{e
1}|^{2}~\cos(\Delta_{A}+\Delta_{S})
\sin\Delta_{A}\sin\Delta_{S}.\label{3nue1}
\end{eqnarray}
and
\begin{eqnarray}
&&P^{\mathrm{IS}}(\bar\nu_{e}\to \bar\nu_{e})=1- 4~|U_{e
3}|^{2}(1-|U_{e3}|^{2})~ \sin^{2}\Delta_{A}
\nonumber\\
&&- 4~|U_{e 2}|^{2}(1-|U_{e 2}|^{2})~ \sin^{2}\Delta_{S}
\nonumber\\
&&-8~|U_{e 3}|^{2}|U_{e
2}|^{2}~\cos(\Delta_{A}+\Delta_{S})
\sin\Delta_{A}\sin\Delta_{S}.\label{3nue2}
\end{eqnarray}
Using the standard parametrization  of the PMNS mixing matrix (\ref{unitmixU1}) from  (\ref{Genexp5}) and  (\ref{Genexp6}) for NS and IS we have respectively
\begin{eqnarray}
&&P^{\mathrm{NS}}(\bar\nu_{e}\to \bar\nu_{e})=1-
\sin^{2}2\theta_{13} \sin^{2}\Delta_{A}
\nonumber\\
&&-
(\sin^{2}2\theta_{12}c^{2}_{13}+\sin^{2}2\theta_{13}c^{4}_{12}) ~ \sin^{2}\Delta_{S}
\nonumber\\
&&-2\sin^{2}2\theta_{13}c^{2}_{12} ~\cos(\Delta_{A}+\Delta_{S}) \sin\Delta_{A}\sin\Delta_{S}.\label{3nue4}
\end{eqnarray}
and
\begin{eqnarray}
&&P^{\mathrm{IS}}(\bar\nu_{e}\to \bar\nu_{e})=1-
\sin^{2}2\theta_{13} \sin^{2}\Delta_{A}
\nonumber\\
&&-
(\sin^{2}2\theta_{12}c^{2}_{13}+\sin^{2}2\theta_{13}s^{4}_{12}) ~ \sin^{2}\Delta_{S}
\nonumber\\
&&-2\sin^{2}2\theta_{13}s^{2}_{12} ~\cos(\Delta_{A}+\Delta_{S})\sin\Delta_{A}\sin\Delta_{S}.\label{3nue5}
\end{eqnarray}
Notice that $P^{\mathrm{IS}}(\bar\nu_{e}\to \bar\nu_{e})$ can be obtained from $P^{\mathrm{NS}}(\bar\nu_{e}\to \bar\nu_{e})$ by the change $c^{2}_{12}\to s^{2}_{12}$.

\subsection{$\nu_{\mu}\to \nu_{e}$ ($\bar\nu_{\mu}\to \bar\nu_{e}$)
 appearance probability}
Vacuum three-neutrino expressions for $\nua{\mu}\to \nua{e}$ transition probabilities are important for analysis of the data of long baseline accelerator experiments  in which matter effects are negligible.
From (\ref{Genexp5}) and (\ref{Genexp6}) we  have
\begin{eqnarray}
&&P^{\mathrm{NS}}(\nua{\mu}\to \nua{e})= 4~|U_{e 3}|^{2}|U_{\mu
3}|^{2}~ \sin^{2}\Delta_{A}
\nonumber\\
&&+4~|U_{e 1}|^{2}|U_{\mu 1}|^{2}~
\sin^{2}\Delta_{S}\nonumber\\
&&-8~\mathrm{Re}~(U_{e 3}U^{*}_{\mu3}U_{e 1}^{*}U_{\mu 1})~\cos(\Delta
_{A}+\Delta _{S}) \sin\Delta
_{A}\sin\Delta _{S}\nonumber\\
&&\mp 8~\mathrm{Im}~(U_{e 3}U^{*}_{\mu 3}U_{e 1}^{*}U_{\mu
1})~\sin(\Delta _{A}+\Delta _{S}) \sin\Delta _{A}\sin\Delta _{S}
.\label{3numue1}
\end{eqnarray}
and
\begin{eqnarray}
&&P^{\mathrm{IS}}(\nua{\mu}\to \nua{e})= 4~|U_{e 3}|^{2}|U_{\mu
3}|^{2}~ \sin^{2}\Delta_{A}
\nonumber\\
&&+4~|U_{e 2}|^{2}|U_{\mu 2}|^{2}~
\sin^{2}\Delta_{S}\nonumber\\
&&-8~\mathrm{Re}~(U_{e 3}U^{*}_{\mu3}U_{e 2}^{*}U_{\mu 2})~\cos(\Delta
_{A}+\Delta _{S}) \sin\Delta
_{A}\sin\Delta _{S}\nonumber\\
&&\pm 8~\mathrm{Im}~(U_{e 3}U^{*}_{\mu 3}U_{e 2}^{*}U_{\mu
2})~\sin(\Delta _{A}+\Delta _{S}) \sin\Delta _{A}\sin\Delta _{S}
.\label{3numue2}
\end{eqnarray}
Using the standard parameterization  of the PMNS mixing
matrix in the case of NS we have
\begin{eqnarray}
&&P^{\mathrm{NS}}(\nua{\mu}\to
\nua{e})=\sin^{2}2\theta_{13}s^{2}_{23}\sin^{2}\Delta_{A}
\nonumber\\&&+(\sin^{2}2\theta_{12}c^{2}_{13}c^{2}_{23}+
\sin^{2}2\theta_{13}c^{4}_{12} s^{2}_{23}+Kc^{2}_{12}\cos\delta)
\sin^{2}\Delta_{S}\nonumber\\&&+(2\sin^{2}2\theta_{13}s^{2}_{23}c^{2}_{12}+K\cos\delta)
~\cos(\Delta_{A}+\Delta_{S}) \sin\Delta_{A}\sin\Delta_{S}\nonumber\\
&&\mp 8 J_{CP}~\sin(\Delta_{A}+\Delta_{S}) \sin\Delta_{A}\sin\Delta_{S}. \label{3numue3}
\end{eqnarray}
Here
\begin{equation}\label{3numue4}
  K=\sin2\theta_{12}\sin2\theta_{13}\sin2\theta_{23}c_{13}.
\end{equation}
and
\begin{equation}\label{Jarskog}
    J_{CP}=\frac{1}{8}K\sin\delta
\end{equation}
is the Jarlskog invariant \cite{Jarlskog:1985ht}.

In the case of the inverted neutrino mass spectrum we find
\begin{eqnarray}
&&P^{\mathrm{IS}}(\nua{\mu}\to
\nua{e})=\sin^{2}2\theta_{13}s^{2}_{23}\sin^{2}\Delta_{A}\nonumber\\&&+(\sin^{2}2\theta_{12}c^{2}_{13} c^{2}_{23}+
\sin^{2}2\theta_{13} s^{4}_{12}s^{2}_{23}-K s^{2}_{12}\cos\delta)
\sin^{2}\Delta_{S} \nonumber\\&&+(2\sin^{2}2\theta_{13}s^{2}_{23}s^{2}_{12}-K\cos\delta)
~\cos(\Delta_{A}+\Delta_{S}) \sin\Delta_{A}\sin\Delta_{S}\nonumber\\
&&\mp 8 J_{CP}~\sin(\Delta_{A}+\Delta_{S}) \sin\Delta_{A}\sin\Delta_{S}. \label{3numue5}
\end{eqnarray}
For the $CP$ asymmetry in the case of  NS (IS) we have
\begin{equation}\label{CPassym}
 A^{CP}_{e\mu} = -16  J_{CP}~\sin(\Delta_{A}+\Delta_{S})
 \sin\Delta_{A}\sin\Delta_{S}.
\end{equation}

\subsection{$\nu_{\mu}\to\nu_{\mu}$ ($\bar\nu_{\mu}\to \bar\nu_{\mu}$) survival probability}

From  (\ref{Genexp4}) for $\nua{\mu}\to\nua{\mu}$ survival probability in the case of the normal and inverted mass ordering we have correspondingly
\begin{eqnarray}
&&P^{\mathrm{NS}}(\nua{\mu}\to \nua{\mu})=1- 4~|U_{\mu
3}|^{2}(1-|U_{\mu3}|^{2})~ \sin^{2}\Delta_{A}
\nonumber\\
&&- 4~|U_{\mu 1}|^{2}(1-|U_{\mu 1}|^{2})~ \sin^{2}\Delta_{S}
\nonumber\\
&&-8~|U_{\mu 3}|^{2}|U_{\mu
1}|^{2}~\cos(\Delta_{A}+\Delta_{S})
\sin\Delta_{A}\sin\Delta_{S}.\label{3mumu}
\end{eqnarray}
and
\begin{eqnarray}
&&P^{\mathrm{IS}}(\nua{\mu}\to \nua{\mu})=1- 4~|U_{\mu
3}|^{2}(1-|U_{\mu 3}|^{2})~ \sin^{2}\Delta_{A}
\nonumber\\
&&- 4~|U_{\mu 2}|^{2}(1-|U_{\mu 2}|^{2})~ \sin^{2}\Delta_{S}
\nonumber\\
&&-8~|U_{\mu 3}|^{2}|U_{\mu
2}|^{2}~\cos(\Delta_{A}+\Delta_{S})
\sin\Delta_{A}\sin\Delta_{S}.\label{3mumu2}
\end{eqnarray}

Using standard parametrization of the PMNS matrix we find
\begin{eqnarray}
&&P^{\mathrm{NS}}(\nua{\mu}\to \nua{\mu})=1- (\sin^{2}2\theta_{23}c^{2}_{13}+\sin^{2}2\theta_{13}s^{4}_{23}) ~ \sin^{2}\Delta_{A}
\nonumber\\
&&-4(c^{2}_{23}s^{2}_{12}+s^{2}_{23}c^{2}_{12}s^{2}_{13} +\frac{K\cos\delta}{4c^{2}_{13}})(1-c^{2}_{23}s^{2}_{12}-s^{2}_{23}c^{2}_{12}s^{2}_{13}-\frac{K\cos\delta}{4c^{2}_{13}})
\sin^{2}\Delta_{S}\nonumber\\
&&-2(\sin^{2}2\theta_{23}c^{2}_{13}s^{2}_{12} + \sin^{2}2\theta_{13}c^{2}_{12}s^{4}_{23} \nonumber\\&&+Ks^{2}_{23}\cos\delta)~\cos(\Delta_{A}+\Delta_{A})\sin\Delta_{A}\sin\Delta_{S}.\label{3mumua}
\end{eqnarray}
and
\begin{eqnarray}
&&P^{\mathrm{IS}}(\nua{\mu}\to \nua{\mu})=1- (\sin^{2}2\theta_{23} c^{2}_{13}+\sin^{2}2\theta_{13}s^{4}_{23}) ~ \sin^{2}\Delta_{A}
\nonumber\\
&&-4(c^{2}_{23}c^{2}_{12}+s^{2}_{23}s^{2}_{12}s^{2}_{13} -\frac{K\cos\delta}{4c^{2}_{13}})(1-c^{2}_{23}c^{2}_{12}-s^{2}_{23}s^{2}_{12}s^{2}_{13}+\frac{K\cos\delta}{4c^{2}_{13}})
\sin^{2}\Delta_{S}\nonumber\\
&&-2(\sin^{2}2\theta_{23}c^{2}_{13}c^{2}_{12} + \sin^{2}2\theta_{13}s^{2}_{12}s^{4}_{23} \nonumber\\&&-Ks^{2}_{23}\cos\delta)~\cos(\Delta_{A}+\Delta_{A})\sin\Delta_{A}\sin\Delta_{S}.
\label{3mumu}
\end{eqnarray}
where $K$ is given by the relation (\ref{3numue4}). Notice that in the case of  long baseline experiments with $\Delta_{A}\simeq 1$ (MINOS, T2K) the term proportional to $\sin^{2}\Delta_{S}$ gives very small contribution to the probability ($\sin^{2}\Delta_{S}\simeq 10^{-3}$).

\subsection{A comment on the possibility to reveal the character of neutrino mass spectrum in future reactor experiments}
Dependence on the neutrino mass ordering of the probability of reactor $\nu_{e}$'s to survive  was  noticed  in the paper \cite{Bilenky:2001jq} in which reactor CHOOZ data were analyzed in the framework of three neutrino mixing. A reactor experiment with reactor-detector distance 20-30 km which could reveal the character of neutrino mass spectrum was proposed in \cite{Petcov:2001sy,Choubey:2003qx}. Later in numerous papers a  possibility to determine  the neutrino mass ordering  in a intermediate  baseline reactor experiment ($\sim$ 50 km) was analyzed in details (see \cite{Li:2013zyd} and references therein). Two reactor experiments JUNO \cite{Li:2014qca} and RENO-50 \cite{Kim:2014rfa}, in which the neutrino mass ordering is planned to be determined, are at preparation at present.

The $\bar\nu_{e}\to \bar\nu_{e}$ survival probability (expressions (\ref{Genexp5}) and  (\ref{Genexp6}) )
can be written in the form
\begin{eqnarray}
&&P(\bar\nu_{e}\to \bar\nu_{e})=1-
\sin^{2}2\theta_{13} \sin^{2}\Delta_{A}
\nonumber\\
&&- 4~X(1-X)~ \sin^{2}\Delta_{S}
\nonumber\\
&&-8\sin^{2}\theta_{13} X~\cos(\Delta_{A}+\Delta_{S})
\sin\Delta_{A}\sin\Delta_{S}.\label{3nue6}
\end{eqnarray}
 In the case of the normal and inverted mass spectra we have respectively
\begin{equation}\label{3nue4}
X=X_{NS}=\cos^{2}\theta_{13}\cos^{2}\theta_{12}
\end{equation}
and
\begin{equation}\label{3nue5}
X=X_{IS}=\cos^{2}\theta_{13}\sin^{2}\theta_{12}
\end{equation}
From the fit of the data that will be obtained  in the reactor JUNO experiment after six years of data taking
the parameters $\Delta m^{2}_{S}$, $\Delta m^{2}_{A}$ and  $\sin^{2}2\theta_{12}$ will be determined with accuracy better than 1\% (see, for example, \cite{He:2013hla}). In the Daya Bay experiment the parameter $\sin^{2}\theta_{13}$ can be determined with accuracy $\sim$ 4\%. Such a precision  will, apparently, allow to distinguish the value
$X\simeq 0.682$ (NS) from the value $X \simeq 0.295$ (IS) (we used best-fit values $\sin^{2}\theta_{12}=0.302,~~\sin^{2}\theta_{13}=0.0227$).

\section{Transitions of flavor neutrinos into sterile states}

Data of atmospheric, solar, reactor and accelerator neutrino oscillation experiments are  described
by the three-neutrino mixing with   two neutrino mass-squared differences $\Delta m_{S}^{2}\simeq 7.5 \cdot 10^{-5}~\mathrm{eV}^{2}$ and $\Delta m_{A}^{2}\simeq 2.4 \cdot 10^{-5}~\mathrm{eV}^{2}$. There exist, however, indications in favor of neutrino oscillations with  mass-squared difference(s) about 1 $\mathrm{eV}^{2}$. These
indications were obtained in following short baseline neutrino experiments (with $L$ ranging from a few meters to about 500 meters):
\begin{enumerate}
  \item  In the LSND experiment \cite{Aguilar:2001ty}. In this experiment neutrinos were produced in decays of $\pi^{+}$'s and $\mu^{+}$'s. Appearance of $\bar\nu_{e}$'s (presumably produced in the transition $\bar\nu_{\mu}\to \bar\nu_{e}$) were detected.
In the MiniBooNE experiment \cite{Aguilar-Arevalo:2013pmq,Conrad:2013mka}. In this experiment an excess of low energy $\nu_{e}$'s  ( $\bar\nu_{e}'s$) was observed.

\item  In the old reactor neutrino experiments. Data of these experiments were reanalyzed in \cite{Mention:2011rk}. In this new analysis  recent calculations of
  the reactor neutrino flux  \cite{Mueller:2011nm,Huber:2011wv} was used.

\item In the calibration experiments, performed with radiative sources by the GALLEX  \cite{Kaether:2010ag} and SAGE \cite{Abdurashitov:2009tn} collaborations. In these experiments a deficit of  $\nu_{e}$'s was observed.
\end{enumerate}
In order to interpret these  data  in terms of neutrino oscillations it necessary to assume that in addition to the flavor  neutrinos $\nu_{e},\nu_{\mu},\nu_{\tau}$ exist also   sterile neutrinos.

Let us consider first 3+1 scheme with three close neutrino masses $m_{i}$ ($i=1,2,3$) and forth mass $m_{4}$
separated from  $m_{i}$  by about $ 1 \mathrm{eV}$ gap. We will choose $p=1$. In the region of $\frac{L}{E}$ sensitive to large neutrino mass-squared difference  ($\frac{\Delta m_{14}^{2}L}{4E}\gtrsim 1$)  we have
$\Delta_{12}\simeq \Delta_{13}\simeq 0$. From  (\ref{Genexp4}) we find in this case
\begin{equation}\label{ster}
 P(\nua{\alpha}\to \nua{\alpha'})= \delta_{\alpha' \alpha }
-4|U_{\alpha 4}|^{2}(\delta_{\alpha' \alpha } - |U_{\alpha' 4}|^{2})\sin^{2}\Delta_{14}.
\end{equation}
From this expression for $\nua{\mu}\to \nua{e}$ appearance probability and $\nua{e}\to \nua{e}$ and $\nua{\mu}\to \nua{\mu}$ disappearance probabilities we have, respectively, the following expressions
\begin{equation}\label{transition2}
P(\nua{\mu}\to \nua{e})=\sin^{2}2\theta_{e\mu}\sin^{2}\Delta_{14},
\end{equation}
\begin{equation}\label{transition3}
P(\nua{e}\to \nua{e})=1-\sin^{2}2\theta_{e e}\sin^{2}\Delta_{14},
\end{equation}
\begin{equation}\label{transition4}
P(\nua{\mu}\to \nua{\mu})=1-\sin^{2}2\theta_{\mu\mu}\sin^{2}\Delta_{14}.
\end{equation}
Here
\begin{equation}\label{transition5}
\sin^{2}2\theta_{e\mu}=4|U_{e 4}|^{2}|U_{\mu 4}|^{2},~\sin^{2}2\theta_{e e}=4|U_{e 4}|^{2}(1-|U_{e 4}|^{2}),~
\sin^{2}2\theta_{\mu\mu}=4|U_{\mu 4}|^{2}(1-|U_{\mu 4}|^{2}).
\end{equation}
Notice that the global  analysis of all  short baseline neutrino data \cite{Giunti:2013aea,Kopp:2013vaa} revealed inconsistency (tension) of existing short baseline data.

Let us consider more complicated 3+2 scheme with 2 masses $m_{4}$  and $m_{5}$ separated from three close masses  $m_{i}$ ($i=1,2,3$) by  about 1 eV gaps. We will choose $p=1$. In the region of $\frac{L}{E}$ sensitive to large neutrino mass-squared differences
$\Delta m^{2}_{14}$ and $\Delta m^{2}_{15}$ we have $\Delta_{12}\simeq \Delta_{13}\simeq 0$. From (\ref{Genexp4}) we find the following expression for  $\nua{l}$ ($l=e,\mu$) survival probability
\begin{eqnarray}
P(\nua{l}\to \nua{l})
&=&1
-4|U_{l 4}|^{2}(1 - |U_{l 4}|^{2})\sin^{2}\Delta_{14}-4|U_{l 5}|^{2}(1 - |U_{l 5}|^{2})\sin^{2}\Delta_{15}\nonumber\\
&+&8~|U_{l 5}|^{2}|U_{l
4}|^{2}\cos(\Delta_{15}-\Delta_{14})\sin\Delta_{15}\sin\Delta_{14}.
\label{Ster5}
\end{eqnarray}
For the  probability of the transitions $\nua{l}\to \nua{l'},~~l'\neq l$ we find
\begin{eqnarray}
P(\nua{l}\to \nua{l'})
= 4|U_{l' 4}|^{2} |U_{l 4}|^{2}\sin^{2}\Delta_{14}
+4|U_{l' 5}|^{2} |U_{l 5}|^{2}\sin^{2}\Delta_{15}
\nonumber\\
+8~\mathrm{Re}~(U_{l' 5}U^{*}_{l 5}U^{*}_{l'
4}U_{l 4})\cos(\Delta_{15}-\Delta_{14})
\sin\Delta_{15}\sin\Delta_{14}\nonumber\\
\pm 8~\mathrm{Im}~(U_{l' 5}U^{*}_{l 5}U^{*}_{l'
4}U_{l 4})\sin(\Delta_{15}-\Delta_{14})\sin\Delta_{15}\sin\Delta_{14}.
\label{Ster4}
\end{eqnarray}
\section{Conclusion}
Discovery of neutrino oscillations is one of the most important recent discovery in the particle physics. After the first stage of investigation of this new phenomenon now with the measurement of the small parameter $\sin^{2}\theta_{13}\simeq 2.5\cdot 10^{-2}$ the era of precision study started. Such fundamental problems of neutrino masses and mixing as
\begin{itemize}
  \item what is the ordering of neutrino masses (normal or inverted),
  \item what is the value of the $CP$ phase $\delta$,
  \item what are precise values (with accuracies better than 1\%) of other oscillation parameters,
\item is the number of massive neutrinos  equal to the number of flavor neutrinos (three) or larger than three
(are sterile neutrinos exist)
\end{itemize}
are planned to be solved by future neutrino oscillation experiments.

At the moment there is no consensus in definition of the large (atmospheric) neutrino mass-squared difference: in different experimental and theoretical papers this parameter is defined differently. Today it is  not so important but with future precision different "atmospheric mass-squared differences" will distinguishable. I believe that universal definition must be accepted.

In this paper for the general case of $n$-neutrino mixing we propose convenient  expression for neutrino transition probability in vacuum in which the unitarity of the mixing matrix is fully employed and freedom of the common phase is used.  As a result only independent quantities (including mass-squared differences) enter into expression for the transition probability.

On the basis of the proposed expression I discuss the problem of the atmospheric neutrino mass-squared difference and comment a possibility to reveal the character of the neutrino mass spectrum in future reactor neutrino experiments.

I thank A. Olshevskiy and C. Giunti for useful discussions.

\end{document}